\documentclass[a4paper]{PoS}
\usepackage[numbers]{natbib}
\usepackage{amsmath}
\usepackage{setspace}
\newcommand{\figheight}{0.22\textheight}

\newcommand{\abovesecspace}{-1.0\baselineskip}
\newcommand{\belowsecspace}{-0.5\baselineskip}

\newcommand{\abovesubsecspace}{-0.5\baselineskip}
\newcommand{\belowsubsecspace}{-0.25\baselineskip}

\newcommand{\vr}{\ensuremath{v_{\rm r}}}

\newcommand{\e}[1]{\ensuremath{\times 10^{#1}}}

\newcommand{\isotope}[2]{\ensuremath{\mathrm {^{#2}#1}}}

\newcommand{\chimera}{{\sc Chimera}}

\newcommand{\TNSE}{\ensuremath{T_{\rm NSE}}}

\newcommand{\Pset}{\ensuremath{\mathbb{P}}}
\newcommand{\Punb}{\ensuremath{\Pset_{\rm unb}}}
\newcommand{\Punbpos}{\ensuremath{\Punb^{+}}}
\newcommand{\Punbneg}{\ensuremath{\Punb^{-}}}
\newcommand{\Punbneghat}{\ensuremath{\lbrack{\Punbneg}\rbrack}}

\newcommand{\Munb}{\ensuremath{M_{\rm unb}}}

\newcommand{\Munbneg}{\ensuremath{\Munb^{-}}}
\newcommand{\Munbneghat}{\ensuremath{\lbrack{\Munbneg}\rbrack}}

\newcommand{\Mposi}{\ensuremath{M^{+}_{\rm i}}}

\newcommand{\eden}[1]{\ensuremath{e_{\rm #1}}}

\newcommand{\edth}{\ensuremath{\eden{th}}}
\newcommand{\edkin}{\ensuremath{\eden{kin}}}
\newcommand{\edgrav}{\ensuremath{\eden{grav}}}

\newcommand{\edtot}{\ensuremath{\eden{tot}}}

\newcommand{\PNS}{proto-NS}

\title{Advancing Nucleosynthesis in Self-consistent, Multidimensional Models of Core-Collapse Supernovae}

\ShortTitle{Nucleosynthesis in Multidimensional Models of CCSNe}

\author{
	\speaker{J. Austin Harris}$^{a}$,
	W. Raphael Hix$^{ba}$,
	Merek A. Chertkow$^{a}$,
	Stephen W. Bruenn$^{c}$,
	
	Eric J. Lentz$^{ab}$,
	O. E. Bronson Messer$^{dba}$,
	Anthony Mezzacappa$^{ae}$,
	John M. Blondin$^{f}$,
	Pedro Marronetti$^{gc}$, and
	Konstantin N. Yakunin$^{abe}$
		\thanks
		{	
			This research was supported by the U.S. Department of Energy Offices of Nuclear Physics; the NASA Astrophysics Theory and Fundamental Physics Program (grants NNH08AH71I and NNH11AQ72I); and the National Science Foundation PetaApps Program (grants  OCI-0749242, OCI-0749204, and OCI-0749248).
			PM is supported by the National Science Foundation through its employee IR/D program. The opinions and conclusions expressed herein are those of the authors and do not represent the National Science Foundation.
			This research was also supported by the NSF  through TeraGrid resources provided by the National Institute for Computational Sciences under grant number TG-MCA08X010; resources of the National Energy Research Scientific Computing Center, supported by the U.S. DoE Office of Science under Contract No. DE-AC02-05CH11231; and an award of computer time from the Innovative and Novel Computational Impact on Theory and Experiment (INCITE) program at the Oak Ridge Leadership Computing Facility, supported by the  U.S. DOE Office of Science under Contract No. DE-AC05-00OR22725.
		}\\
    \llap{$^a$}
    	Department of Physics and Astronomy, University of Tennessee,
	    Knoxville, TN 37996, USA\\
	\llap{$^b$}
	    Physics Division, Oak Ridge National Laboratory,
	    Oak Ridge, TN 37831, USA\\
	\llap{$^c$}
		Department of Physics, Florida Atlantic University,
		Boca Raton, FL 33431, USA\\
	\llap{$^d$}
		National Center for Computational Sciences, ORNL,
		Oak Ridge, TN 37831, USA\\
	\llap{$^e$}
		Joint Institute for Computational Sciences, ORNL,
		Oak Ridge, TN, 37831 USA\\
	\llap{$^f$}
		Department of Physics, North Carolina State University,
		Raleigh, NC 27695, USA\\
	\llap{$^g$}
		Physics Division, National Science Foundation,
		Arlington, VA 22230, USA\\
    E-mail: \email{jharr100@utk.edu}, \email{raph@ornl.gov}
}

\abstract{
We investigate core-collapse supernova (CCSN) nucleosynthesis in polar axisymmetric simulations using the multidimensional radiation hydrodynamics code {\sc Chimera}.
Computational costs have traditionally constrained the evolution of the nuclear composition in CCSN models to, at best, a 14-species $\alpha$-network.
Such a simplified network limits the ability to accurately evolve detailed composition, neutronization and the nuclear energy generation rate.
Lagrangian tracer particles are commonly used to extend the nuclear network evolution by incorporating more realistic networks in post-processing nucleosynthesis calculations.
Limitations such as poor spatial resolution of the tracer particles, estimation of the expansion timescales, and determination of the ``mass-cut'' at the end of the simulation impose uncertainties inherent to this approach.
We present a detailed analysis of the impact of these uncertainties on post-processing nucleosynthesis calculations and implications for future models.
}

\FullConference{XIII Nuclei in the Cosmos\\
                 7-11 July, 2014\\
                 Debrecen, Hungary}

\begin{document}
{\setstretch{0.89}

\section{Introduction}
\label{sec:intro}
\vspace{\belowsecspace}

The deaths of massive stars ($M>$ 8--10~$M_\odot$) as core-collapse supernovae (CCSNe) are an important link in our chain of origins from the Big Bang to the present.
They are the dominant source of elements in the periodic table between oxygen and iron \citep{WoWe95,ThNoHa96}, and there is growing evidence they or some related phenomenon are indeed responsible for producing half the elements heavier than iron \citep{ArSaTh04}.

Recently, multi-dimensional simulations by several groups utilizing spectral neutrino transport have successfully produced explosions for a variety of progenitors \citep{BuJaRa06,MaJa09,MuJaHe12,BrMeHi13,BrLeHi14} though delayed by hundreds of milliseconds compared to their non-spectral counterparts \citep{HeBeHi94,FrWa04}.
The success of these models is in part due to computational prowess associated with multidimensional explosions and a deeper understanding of the large scale convective behavior.
This understanding includes the nature of the hydrodynamic instability of the stalled bounce shock to non-radial perturbations favoring low order modes, termed the Stalled Accretion Shock Instability \citep[SASI;][]{BlMeDe03}.

Though not the driver of CCSNe, an accurate depiction of the thermonuclear reaction network is crucial in understanding the resulting nucleosynthesis and galactic chemical evolution.
Unfortunately, complications with the complexity of self-consistent models and their past failures to produce explosions led to a tendency to model nucleosynthesis from CCSNe using a parameterized kinetic energy piston \citep{RaHeHo02,UmNo08} or thermal energy bomb \citep{NaShSa98}, often in spherical symmetry.
These parameterizations fall short in the inner region of the star, where interactions with the intense neutrino flux from the \PNS\ are important \citep{FrYoBe08}.
More recent simulations utilizing spectral neutrino transport \citep{RaJa02,BuRaJa06,MaJa09,BrLeHi14} witness a decrease in the neutronization in the outer part of the neutrino reheating region as a result of these interactions, impacting the nuclear composition of the ejecta.
Despite the profusion of exploding first-principles models, very few have been continued sufficiently long after bounce to characterize the ejecta nucleosynthesis.

In this paper, we critically analyze the methods used to study the nucleosynthesis in the four two-dimensional models of \citep{BrLeHi14} evolved with the \chimera\ code by examining how some uncertainties stemming from ongoing hydrodynamic activity at 1--2~s after bounce can impact post-processing nucleosynthesis results.

\chimera\ is a multidimensional radiation hydrodynamics code for stellar core collapse with five principal components: hydrodynamics, neutrino transport, self-gravity, a nuclear reaction network, and a nuclear equation of state (see \citep{BrLeHi14} for details).
We track the composition in NSE regions using either 4-species, neutrons, protons, $\alpha$-particles and a representative heavy nucleus, or 17-species, depending on the electron fraction.
For regions not in NSE, the nucleosynthesis is computed within the constraints of an $\alpha$-network ($\alpha$, \isotope{C}{12}-\isotope{Zn}{60}) by XNet, a fully implicit thermonuclear reaction network code \citep{HiTh99b}.

\vspace{\abovesecspace}
\section{Nucleosynthesis Uncertainties}
\label{sec:uncertainties}
\vspace{\belowsecspace}

A number of factors contribute to uncertainties in CCSN nucleosynthesis.
Here we discuss how an unresolved ``mass-cut'', parameterizations of thermodynamic trajectories, and tracer particle resolution can affect the composition of the ejecta.

\vspace{\abovesubsecspace}
\subsection{Determination of the ``mass-cut''}
\label{sec:masscut}
\vspace{\belowsubsecspace}

Core-collapse supernovae are highly asymmetric events driven by complex and/or turbulent flows.
The implications of multidimensionality on the nucleosynthesis are lost in 1D simulations where a clear distinction, the mass-cut, is made between matter which is ejected to the interstellar medium and that which falls onto the \PNS.
Extending this distinction to 2D and 3D simulations is an arduous task which requires evolving a model well beyond the initial development of an explosion until downflows which have long been cut-off from the rest of the star at the shock cease accreting onto the \PNS.
As opposed to the typical parameterized models, the computational cost of running simulations with spectral neutrino transport limits our ability to fully resolve this multi-dimensional ``mass-cut'' with \chimera\, despite 1-2~s of evolution after bounce.

Following the treatment of the explosion energy in \citep{BrLeHi14}, we use the total energy density, $\edtot = \edkin + \edth + \edgrav$, to define the unbound ejecta to be represented by particles for which $\edtot > 0$.
We label this set of unbound particles as \Punb\ and the corresponding mass as $\Munb = 0.41~M_\odot$ in B12-WH07.
The total mass of the ejecta, not to be confused with \Munb, also includes an additional $9.01~M_\odot$ as yet unshocked matter.
Ideally, \Punb\ would have a one-to-one correspondence with the ejecta observed in supernovae.
However, due to the work required to lift the stellar envelope out of the star's gravitational well, \Punb\ is an over-estimate of particles which would ultimately be ejected.
Furthermore, some particles in \Punb\ have negative radial velocities and cannot be reliably extrapolated for post-processing nucleosynthesis.
For this reason, it is helpful to divide \Punb\ by radial velocity, which we label as \Punbpos\ and \Punbneg\ for $\vr > 0$ and $\vr < 0$ respectively.
We find that the ultimate ``fates'' of particles in \Punbneg\ as either ejecta or part of the \PNS\ are often unknowable at the end of the simulation.
Consequently, the mass represented by \Punbneg, \Munbneg, is one indication of uncertainty in the total ejecta mass.
The length of time we must evolve a model in order to keep this uncertainty manageable is characterized by trends in $\Munbneg(t)$.

\begin{figure}[t]
\begin{center}
\includegraphics[height=\figheight]{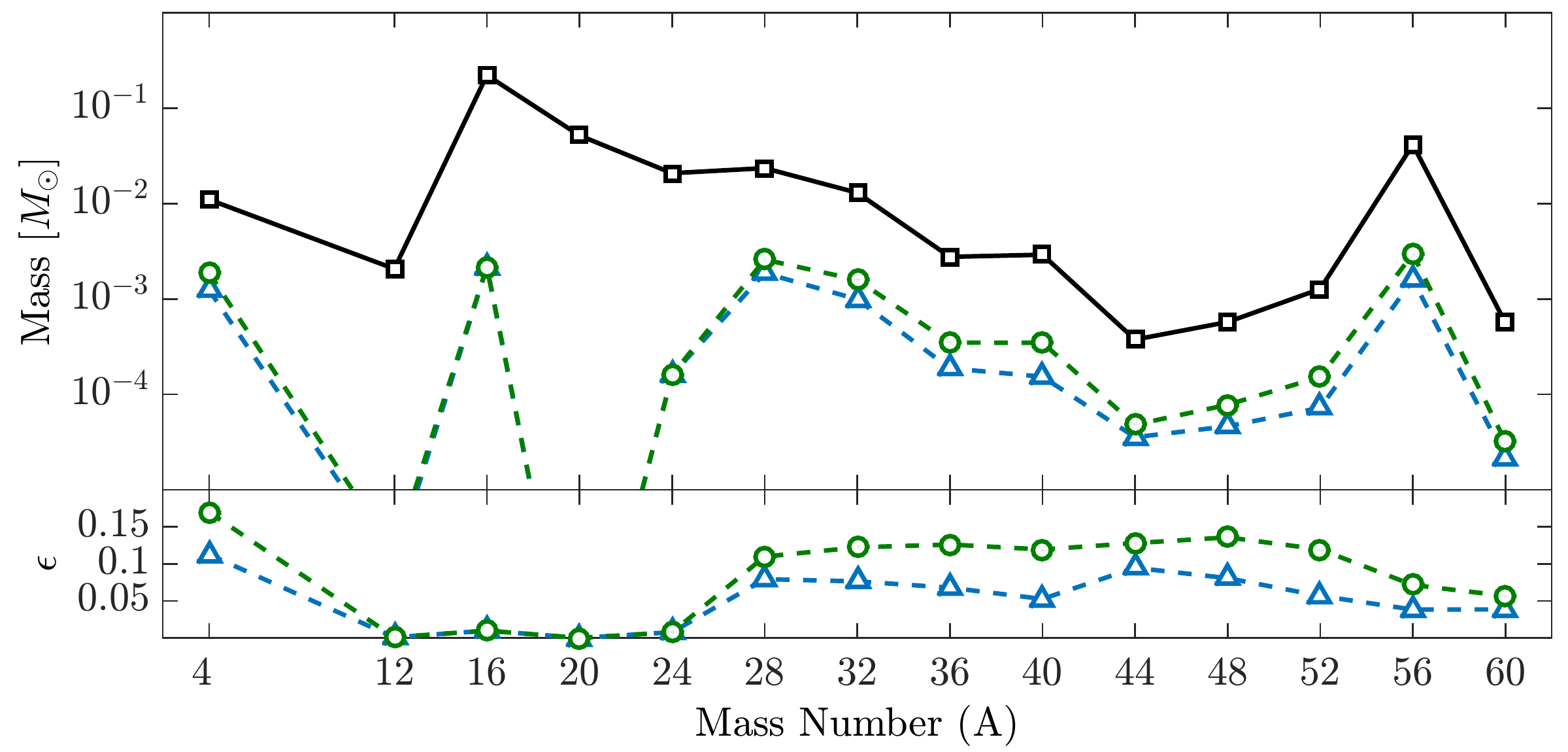}
\end{center}
\caption{\label{fig:b12masscut}
(a) Total unbound mass (black line) and (b) fraction thereof, $\epsilon$, represented by \Punbneg\ (blue line) and \Punbneghat\ (green line).
Post-processing nucleosynthesis calculated using the $\alpha$-network and $\TNSE = 8~GK$.
}
\end{figure}

For the B12-WH07 model, \Munbneg\ shows relatively little change over the last 100~ms of the simulation.
However, closer inspection reveals that while the number of particles in \Punbneg\ mirrors the behavior of \Munbneg, the individual particles in \Punbneg\ are changing as convective flows move particles between \Punbneg\ and \Punbpos.
Therefore, we identify all particles in \Punbneg\ at any time in the last 100~ms of the simulation, which we label \Punbneghat, as having indeterminate ``fates'' and contributing to a better estimated uncertainty in the ejecta mass.
For B12-WH07, $\Munbneg = 9.15\e{-3}~M_\odot$ at the end of the simulation (1.41~s after bounce), but $\Munbneghat = 1.66\e{-2}~M_\odot$.

The impact of this uncertainty on the nucleosynthesis products in B12-WH07 is mostly confined to ($A \geq 28$) for both \Punbneg\ and \Punbneghat\ (see Figure~\ref{fig:b12masscut}).
The effect on the production of these species can be understood by considering the region of the star where the ``mass-cut'' has not yet been determined.
For B12-WH07, this region is confined to the inner 5,000~km of the star around a cut-off downflow rich in \isotope{Si}{28} which continues to accrete onto the \PNS\ long after the development of the explosion.
One would then expect \isotope{Si}{28} and the products of Si-burning to be most impacted by uncertainty in the identification of the ejecta, since the star is too cool at this point for nuclear disassociation.




\vspace{-0.5\baselineskip}
\subsection{Thermodynamic extrapolation}
\label{sec:extrap}
\vspace{-0.25\baselineskip}

\begin{figure}[t]
\begin{center}
\includegraphics[height=\figheight]{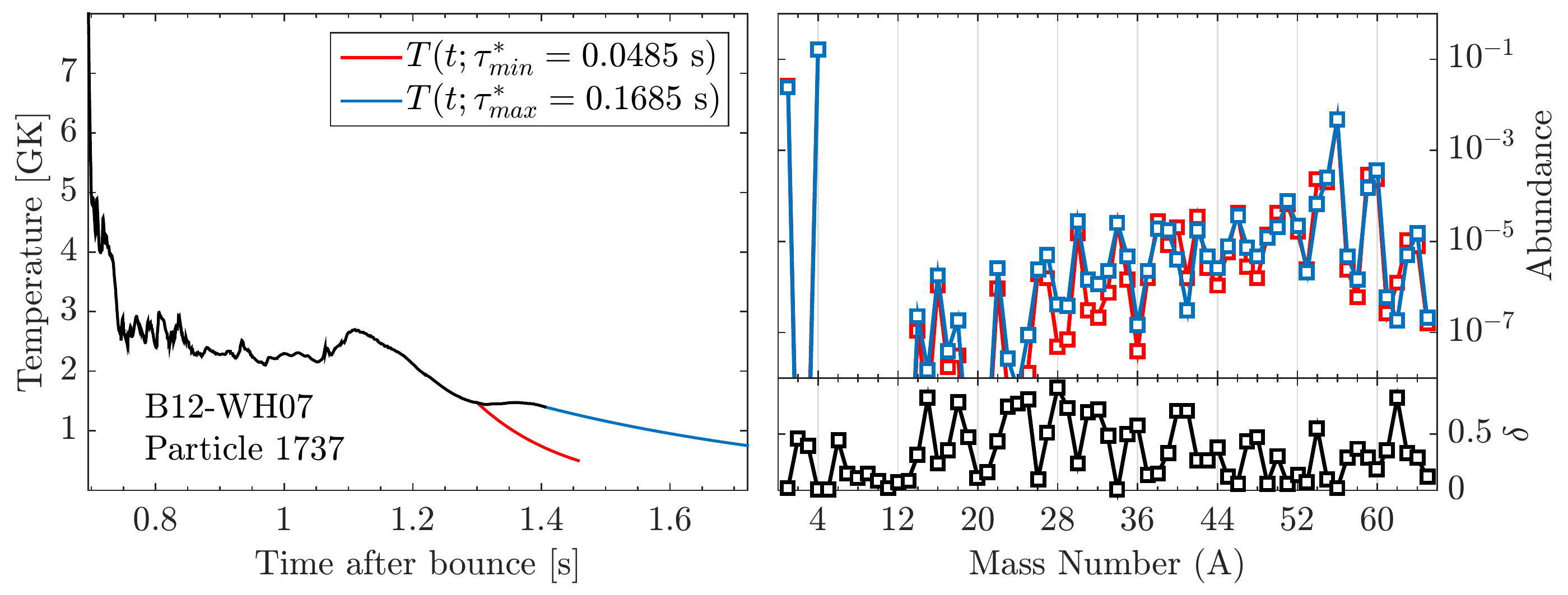}
\end{center}
\caption{\label{fig:p1737extrap}
(a) Thermodynamic extrapolations to 0.5~GK and (b) final abundance profiles for a single particle (\#1737) calculated using ${\tau^{*}_{max}}$ (blue line) and ${\tau^{*}_{min}}$ (red line) ($Y_{65} = \sum\nolimits_{A > 64} Y_i$).
(c) Relative deviation of the abundances between the two extrapolations plotted for each species $i$ as $\delta_i = |log_{10}(Y(\tau^{*}_{max})/Y(\tau^{*}_{min}))|$.
}
\end{figure}

At 1.41~s after bounce in B12-WH07, with temperatures below 3~GK for all particles in \Punbpos, the nuclear reactions which account for the bulk of the nickel production in CCSN have ceased.
However, secondary nuclear burning processes will continue to alter the abundance distribution until the matter freezes out \citep{WoMaWi94}, and neutrino-induced reactions will continue until the temperature of the ejecta falls below $\approx$0.5~GK \citep{FrHaLi06}.
To extend our models beyond this point in time, we extrapolate the thermodynamic conditions under the common assumption of isentropic expansion (i.e. $T^3/\rho = constant$) \citep{HoFo60}.
We estimate an expansion timescale, $\tau^{*}$, by averaging the instantaneous expansion timescale, $\tau = -\rho/\dot{\rho}$, during expanding time periods over either the final 150~ms of the simulation or until $T^3/\rho$ deviates by more than 5\% from the final value.

Of course, any extrapolation will fail to capture future hydrodynamic activity which deviates from true isentropic expansion.
The extent to which this can affect the estimation of the expansion timescale, and consequently the final abundance distribution, is demonstrated in an extreme case for one tracer particle (\#1737) in Figure~\ref{fig:p1737extrap}.
In order to determine the significance of uncertainties in the thermodynamic extrapolation, we post-process the thermodynamic profiles generated by extrapolating from points in time in the last 150~ms of the simulation corresponding to the minimum and maximum estimates of the expansion timescale ($\tau^{*}_{min}$ and $\tau^{*}_{max}$ respectively).

A faulty assumption of isentropic expansion from the end of the simulation can lead to non-trivial uncertainties in the composition of individual particles (see Figure~\ref{fig:p1737extrap}).
The extent to which this impacts the total ejecta mass for each nuclear species $i$, \Mposi, represented by \Punbpos\ (the particles for which we perform extrapolations) is shown in Figure~\ref{fig:b12ejectamassextrap}.
Due to the relatively low temperatures of particles in \Punbpos\ during the final 150~ms of this simulation, the nuclear products which are particularly susceptible to extrapolation uncertainties are those with either a sensitivity to the timescale of \textit{$\alpha$-rich freeze-out} (e.g. \isotope{Ti}{44},...) or dependencies on neutrino-induced reaction pathways ($A > 64$).

\begin{figure}[h]
\begin{center}
\includegraphics[height=\figheight]{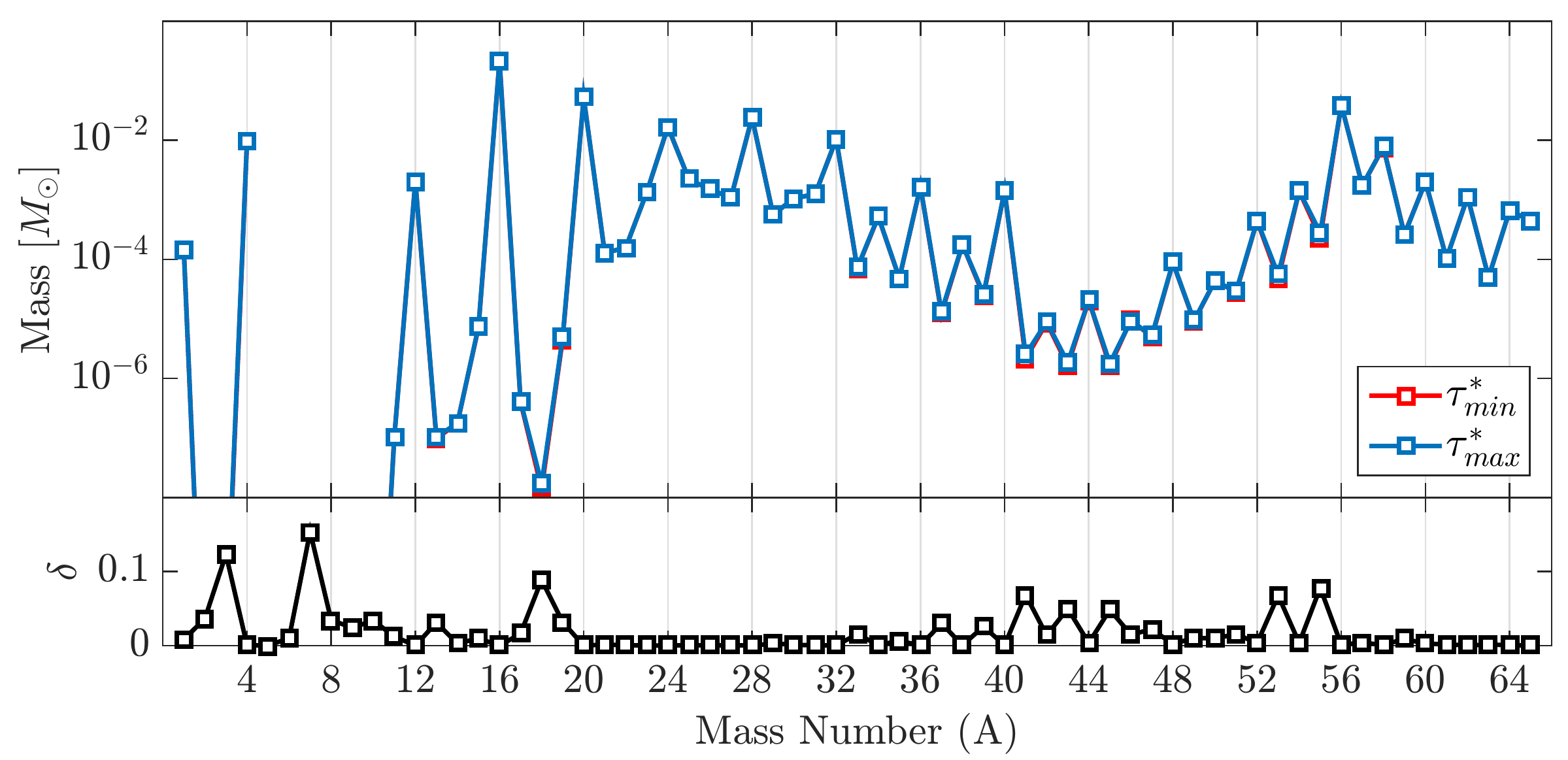}
\end{center}
\caption{\label{fig:b12ejectamassextrap}
(a) \Mposi\ at 0.5~GK and calculated for expansion timescales ${\tau^{*}_{max}}$ (blue line) and ${\tau^{*}_{min}}$ (red line) for each particle in \Punbpos\ ($M^{+}_{65} = \sum\nolimits_{A > 64} \Mposi$).
(b) Relative deviation of the composition between the two extrapolations plotted for each species $i$ as $\delta_i = |log_{10}(\Mposi(\tau^{*}_{max})/\Mposi(\tau^{*}_{min}))|$.
The nucleosynthesis calculation uses a 1160-species nuclear network with neutrino interactions capable of tracking the $\nu p$-process \citep{FrHaLi06}.
}
\end{figure}

\vspace{-0.5\baselineskip}
\subsection{Particle resolution}
\label{sec:resolution}
\vspace{-0.25\baselineskip}

Tracer particles in \citep{BrLeHi14} are initially distributed into ``rows'' at equally spaced mass shells beginning 0.1~$M_\odot$ inside the edge of the iron core.
The particles within each ``row'' are similarly placed such that they represent uniform volume: $\Delta(cos \theta) = \pi/N$, where $N = 40$ particles per mass shell for the models in \citep{BrLeHi14}.
For B12-WH07, this translates to 4,000 particles, each representing $\approx 1.87\e{-4}~M_\odot$, extending from $\approx$890~km to the carbon-enriched oxygen-shell at $\approx$15,000~km.

To quantify the effect of particle resolution on the nucleosynthesis, we post-process thermodynamic profiles from the tracer particles which are generated such that they replicate the conditions of nuclear burning inside \chimera.
This is done by performing the post-processing calculation without the extrapolation described in the previous section using the same network used in the simulation.
Furthermore, the initial conditions for nuclear burning are determined by generating a NSE composition at the point in time matching the transition which occurs \textit{in situ}.

The total ejected mass of each nuclear species $i$ from \chimera\ \textit{in situ} nucleosynthesis, $M^{\chimera}_i$ is calculated by integrating over all zones where $\edtot > 0$.
We perform the equivalent post-processing calculation, applying the $\edtot > 0$ criteria to particles instead of zones, and compare the resulting mass of nucleosynthesis products, $M^{PP}_i$, in Figure~\ref{fig:b12commutator}.
There is general agreement between these two methods for most species.
However, there is a stark discrepancy in the total mass of \isotope{Ti}{44}, with the value from \textit{in situ} calculations, $M^{\chimera}_{Ti-44} \approxeq 1.08\e{-3}~M_\odot$ being greater than that from post-processing, $M^{PP}_{Ti-44} \approxeq 1.24\e{-4}~M_\odot$, by nearly an order of magnitude.
There is also disagreement, albeit less severe, in the \isotope{Cr}{48} mass.
To understand the origin of this inconsistency, consider that the isotopes affected are products of \textit{$\alpha$-rich freeze-out} occurring in low-density, expanding ejecta.
Furthermore, remember that particles are initially distributed such that each particle represents an equal mass and are therefore susceptible to poor spatial resolution at low densities.

As a test of the NSE transition criteria employed in \chimera, we compare post-processing results using the $\alpha$-network and vary the conditions at which the transition to nuclear burning from the NSE composition occurs (Figure~\ref{fig:b12commutator}).
The \textit{in situ} calculation transitions out of NSE at a temperature $\approx$2--3~GK lower than the tested value of 8~GK.
As a result, we see a significant shift in the masses of \isotope{He}{4}, \isotope{Ti}{44}, \isotope{Cr}{48}, and \isotope{Zn}{60}.
This argues for a stricter set of criteria for the assumption of NSE than the $\approx$5-6~GK commonly used within models of the CCSN mechanism, so that one may properly track nuclear burning throughout freeze-out.

\begin{figure}[t]
\begin{center}
\includegraphics[height=\figheight]{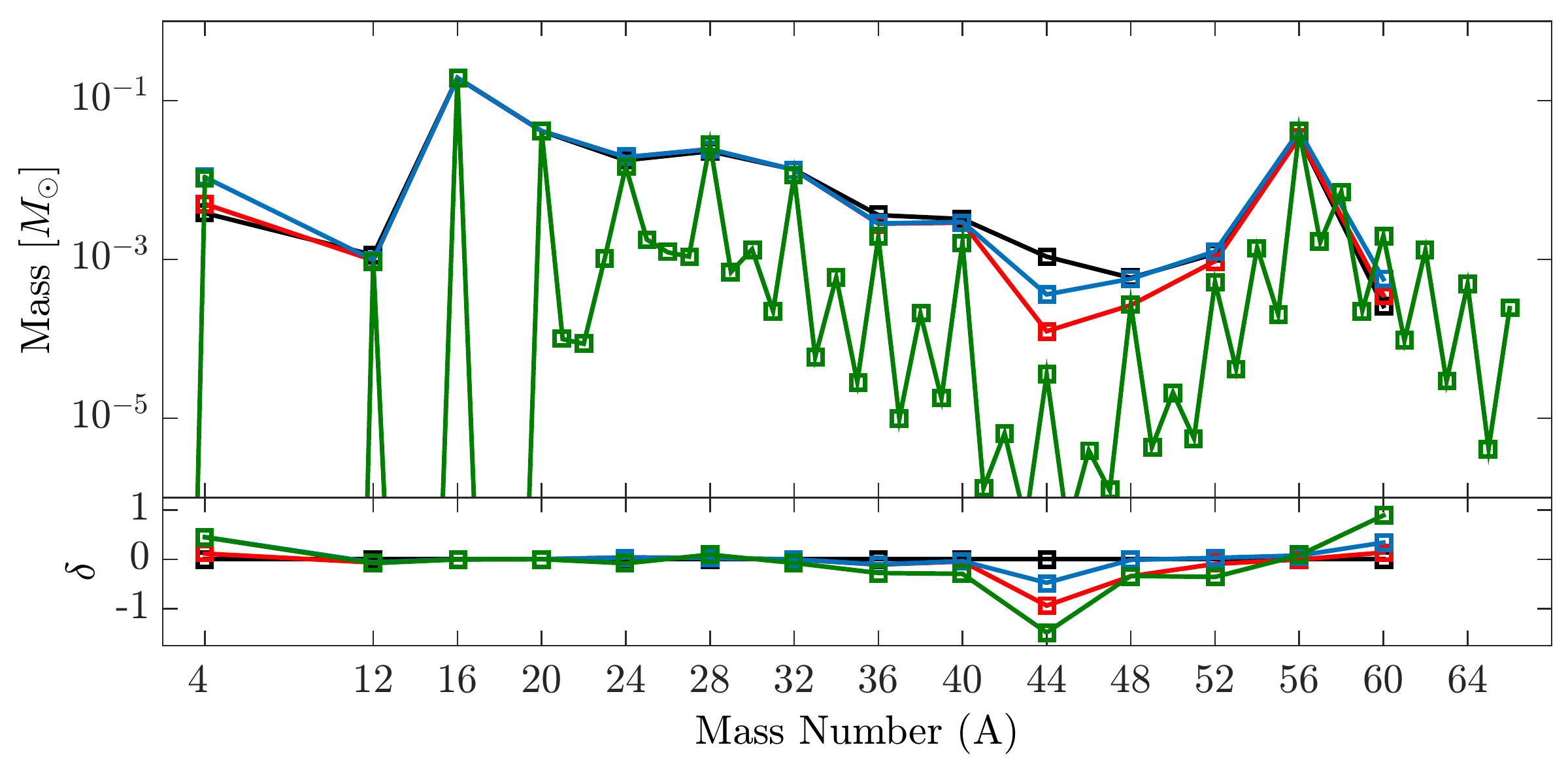}
\end{center}
\caption{\label{fig:b12commutator}
(a) $M^{\chimera}_i$ (black line) and $M^{PP}_i$ at 1.2~s after bounce using different values for \TNSE:
\TNSE\ as a function of density, identical to that used in \chimera\ with the same $\alpha$-network used in the simulation (red line);
$\TNSE = 8~GK$ evolved with the $\alpha$-network (blue line) and with the 150-species network (green line).
(b) Deviations from the original \textit{in situ} nucleosynthesis plotted for each species $i$ as $\delta_i = log_{10}(M^{PP}_i/M^{\chimera}_i)$.
}
\end{figure}

\vspace{\abovesecspace}
\section{Summary}
\label{sec:summary}
\vspace{\belowsecspace}

In this paper, we have discussed some of the uncertainties which complicate post-processing nucleosynthesis calculations from \textit{ab initio} multi-dimensional CCSN models, using the 12~$M_\odot$ model of \citep{BrLeHi14} as an example.
A more detailed analysis of these uncertainties and a broader discussion of nucleosynthesis in \citep{BrLeHi14} is forthcoming \citep{HaHiLe14,HaHiLe15}.
Even after $\approx$1.41~s of post-bounce evolution, the multi-dimensional ``mass-cut'' remains unresolved, where it may impact the production of nuclear species in borderline ejecta near ongoing accretion flows.
In B12-WH07, the effect of the indeterminate mass-cut is most prevalent for $A \geq 28$ and represents <15\% of the total unbound mass represented by particles for any particular isotope, but the sensitivity for individual species is dependent on the composition of the cut-off downflow(s).
Also, despite temperatures below 3~GK for all of the unbound particles, local hydrodynamic deviations from isentropic expansion continue to play a non-trivial role on secondary nuclear processes by altering the expansion timescale estimate necessary for extrapolation to freeze-out at $\approx$0.5~GK.

Both the indeterminate ``mass-cut'' and expansion timescale uncertainties could, in theory, be reduced by extending the simulation.
However, given the inadequate spatial resolution of the tracer particles, we cannot rely entirely on post-processing methods to obtain an accurate representation of the nucleosynthesis.
Improving upon the existing \textit{in situ} $\alpha$-network with a more realistic 150-species nuclear network capable of properly tracking neutronization and energy released via particle captures is an important step towards resolving this problem.
However, the availability of computational resources constrains our ability to incorporate a larger network in all of our models.
We estimate the impact a larger, \textit{in situ} network may have in Figure~\ref{fig:b12commutator} by post-processing the nucleosynthesis calculation using the 150-species network and comparing the resulting composition to that obtained using the $\alpha$-network.
There are significant differences in the total unbound mass for $A \geq 36$ with the exception of $A = 56$.
These differences can be largely attributed to the availability of additional reaction pathways during explosive burning, particularly those involving $(n,\gamma)$ and $(p,\gamma)$ reactions \citep{HiTh96,TiHoWo00}.
Without an \textit{in situ} large network simulation, we are unable to quantify how spatial resolution of the tracer particles may impact nucleosynthesis with realistic nuclear networks.


\vspace{\abovesecspace}


}
\end{document}